\documentclass[slac_one]{revtex4}
\usepackage{graphicx}
\usepackage{fancyhdr}
\newcommand{\BR}{{\cal B}}

\newcommand{\psp}{\psi^{\prime}}

\newcommand{\jpsi}{J/\psi}

\newcommand{\x}{X(3872)}
\newcommand{\EE}{e^+e^-}

\newcommand{\pp}{\pi^+\pi^-}

\newcommand{\bbb}{B\bar{B}}

\newcommand{\ppjpsi}{\pi^+\pi^- J/\psi}

\newcommand{\kkpi}{K_S^0K^+\pi^- + c.c.}
\newcommand{\infb}{\rm fb^{-1}}
\newcommand{\gev}{\rm GeV}
\newcommand{\mev}{\rm MeV}

\newcommand{\ev}{\rm eV}
\newcommand{\gevcs}{{\gev}/c^2}
\newcommand{\mevcs}{{\mev}/c^2}

\newcommand{\beq}{\begin{equation}}
\newcommand{\eeq}{\end{equation}}
\newcommand{\bitm}{\begin{itemize}}
\newcommand{\eitm}{\end{itemize}}

\def\Journal#1#2#3#4{{#1} {\bf #2} (#4) #3}

\def\PRL{Phys. Rev. Lett.}
\def\PRD{Phys. Rev. D}

\def\EPJC{Eur. Phys. J. C}

\begin{document}

\title{New and conventional charmonium states} 

%

\author{Xiaolong Wang}
\affiliation{Institute of High Energy Physics, Chinese Academy of Sciences, Beijing 100049, China}


\begin{abstract}
There are many exotic properties of the charmonium and charmoniumlike states above $D\bar{D}$ threshold.
The recent experimental results from BaBar and Belle on the new and conventional charmonium states are reviewed in this talk.

\end{abstract}

\maketitle

\thispagestyle{fancy}


\section{Introduction}

Both charmonium and bottomonium are proven success stories
of QCD~\cite{review}.
Below the $D\bar{D}$ threshold, all charmonia have been established,
the most recent being the  $h_c$~\cite{hc}, and there is good agreement between experimental
mass measurements and predictions based upon potential models.

There are many charmonium and  charmoniumlike states observed above the $D\bar{D}$ threshold in the past
ten years. Some of these are good candidates for the charmonia predicted
in different models. On the other hand, many have unusual quantum numbers, which may indicate that exotic states such as
multi-quark state, molecule, hybrid, or glueball, have been
observed~\cite{review}.

In this talk, the new and conventional charmonium states at BaBar and Belle are reviewed.

\section{$X(3872)$ and possible partners}

\begin{figure*}[htbp]
\centering
\includegraphics[height=4.1cm,width=4.1cm]{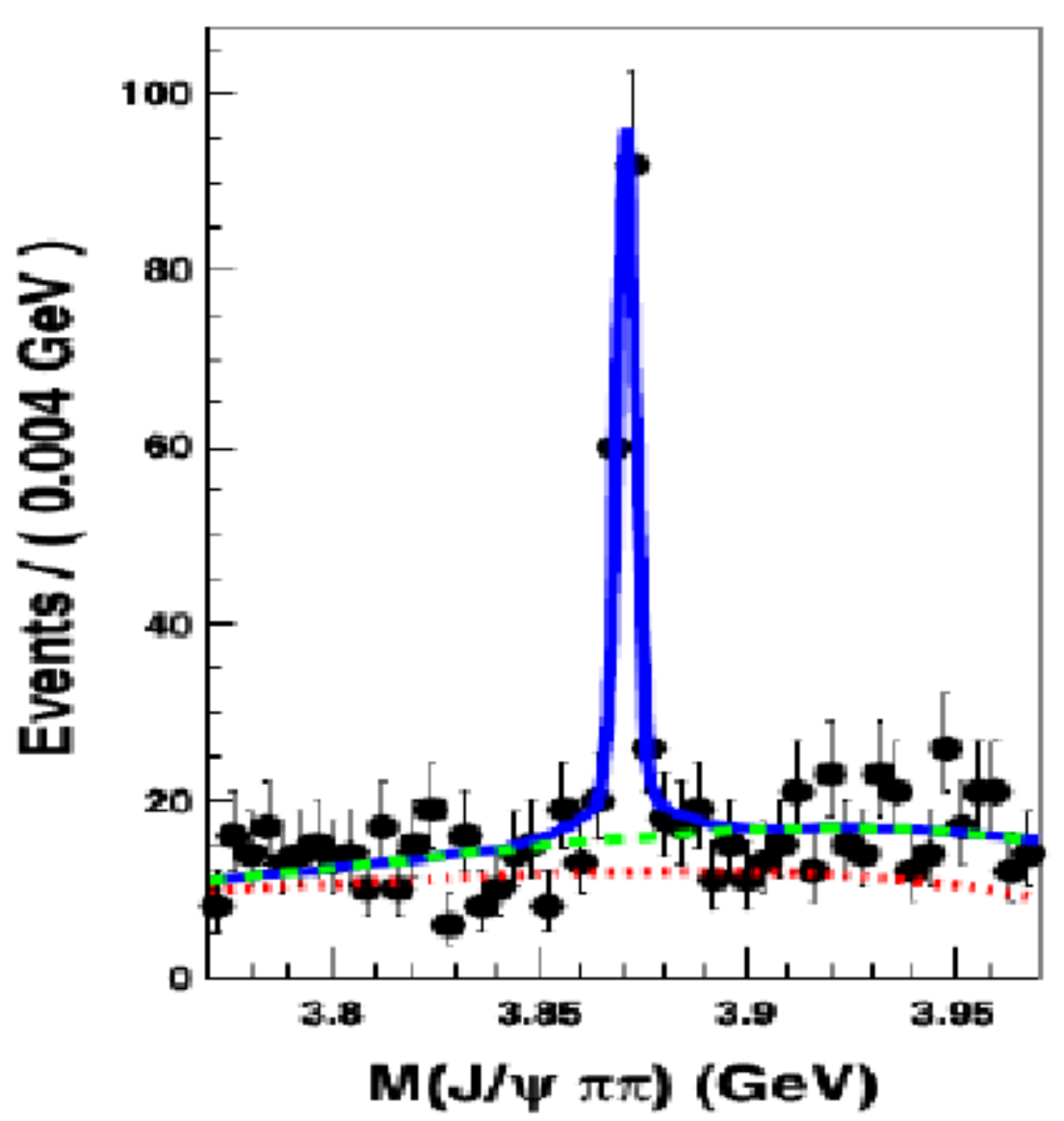}
\includegraphics[height=4.1cm,width=4.1cm]{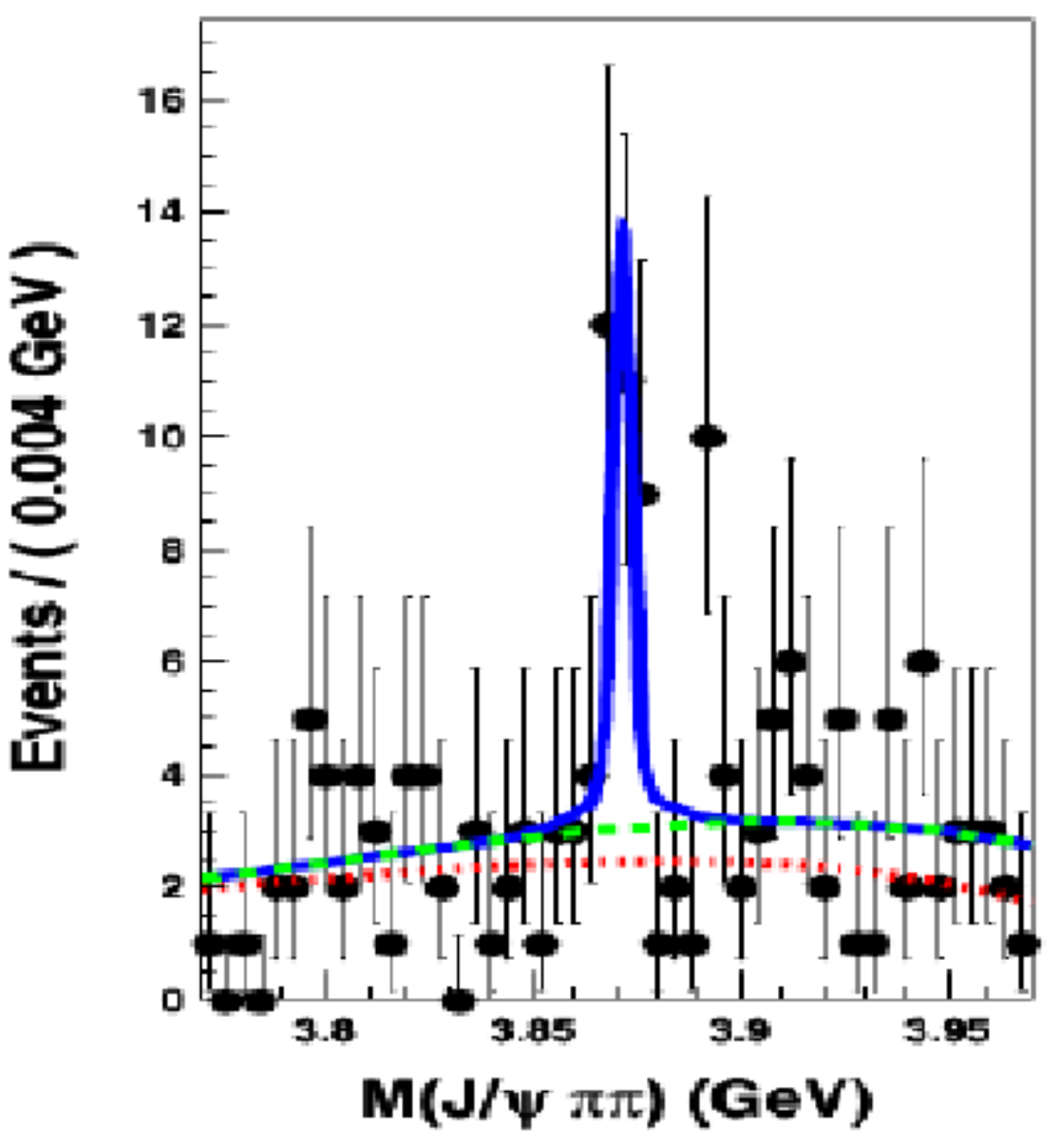}
\caption{The $X(3872)\to\ppjpsi$ signals in $B^{\pm}$ (left) and
$B^0$ (right) decay modes.} \label{x3872-update}
\end{figure*}

$X(3872)$ was discovered by Belle in the decay mode $B\to K+ X(3872)(\to\ppjpsi)$ in 2003~\cite{x3872-1}.
Recently, Belle uses its full data sample, which contains
$772\times 10^6 \bbb$, to update the results on $X(3872)$~\cite{x3872-2}. Figure~\ref{x3872-update} shows the $X(3872)$
signals from $B$ decays. The mass is measured to be $3871.84\pm0.27\pm0.19~\mevcs$,
and the width is determined to be $\Gamma_{X(3872)} < 1.2~\mev$ at 90\% C.L. The mass difference from $B^+$ and
$B^0$ decays is $\Delta M_{\x} = -0.69\pm0.97\pm0.19~\mevcs$, which is consistent with zero.
The angular analysis shows that both $J^{PC} = 1^{++}$ and $J^{PC} = 2^{-+}$ hypotheses describe the data well.

Since a charged partner could be possible if $\x$ is exotic, Belle performs a search for a possible partner
in $\x^+\to\jpsi\rho (\to\pi^+\pi^0)$. There is no signal in $B^+$ decays nor in $B^0$ decays, and the upper limits
of branching fractions are determined to be $\BR(\overline{B}^0\to X^+ K^-)\times\BR(X^+\to\jpsi\rho^+)<4.2\times 10^{-6}$
and $\BR(B^+\to X^+ K^0)\times\BR(X^+\to\jpsi\rho^+)<6.1\times 10^{-6}$.

\begin{figure*}[t]
\centering
\includegraphics[height=4.1cm]{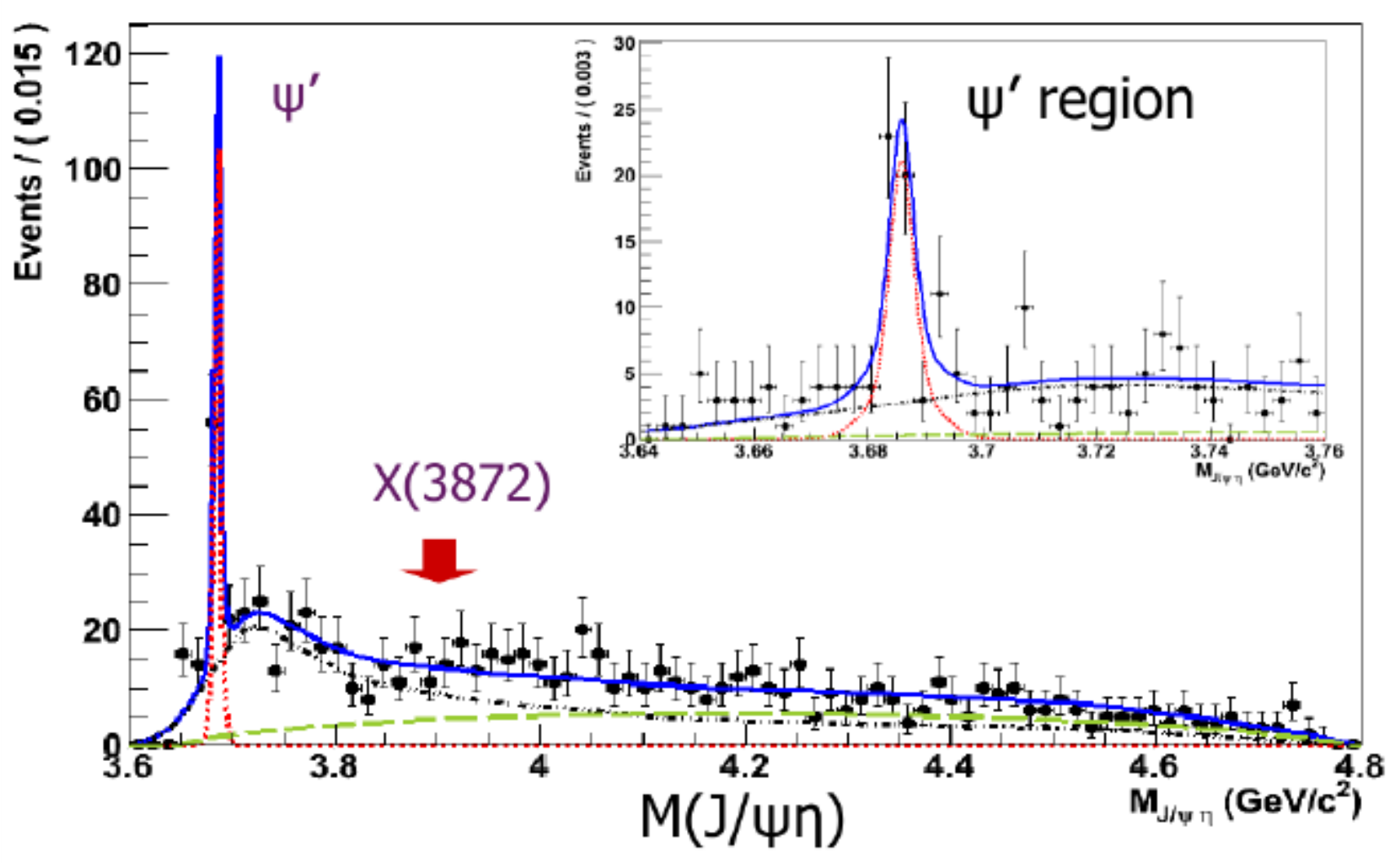}
\caption{The $M_{\eta\jpsi}$ distribution in $B$ decays from Belle. Except for the expected $\psp$ signal, there is no
obvious signal in the $\eta\jpsi$ final state.} \label{x-etajpsi}
\end{figure*}

 A possible $C-$odd neutral partner of $\x$ is also searched for in $\eta\jpsi$ and $\gamma\chi_{c1}$ final states. BaBar searched in
the $\eta\jpsi$ final state with $90\times 10^6~\bbb$ data sample soon after the discovery of $\x$ but found no signal.
Belle performs a similar search
with the full data sample. Figure~\ref{x-etajpsi} shows the resulting $M_{\eta\jpsi}$ distribution. There is still no
obvious $C-$odd partner; Belle sets an upper limit
$\BR(B^+\to\x K^+)\times\BR(X\to\eta\jpsi)<3.8\times 10^{-6}$ at 90\% C.L.

In Belle's search in the $\gamma\chi_{c1}$ final state, there is also no evidence for the neutral $C-$odd $X$ state. The upper limit is
$\BR(B^+\to\x K^+)\times \BR(X\to\gamma\chi_{c1}) < 2.0\times 10^{-6}$ at 90\% C.L.

\section{First evidence of $\psi_2$}

\begin{figure*}[htbp]
\centering
\includegraphics[height=4.1cm]{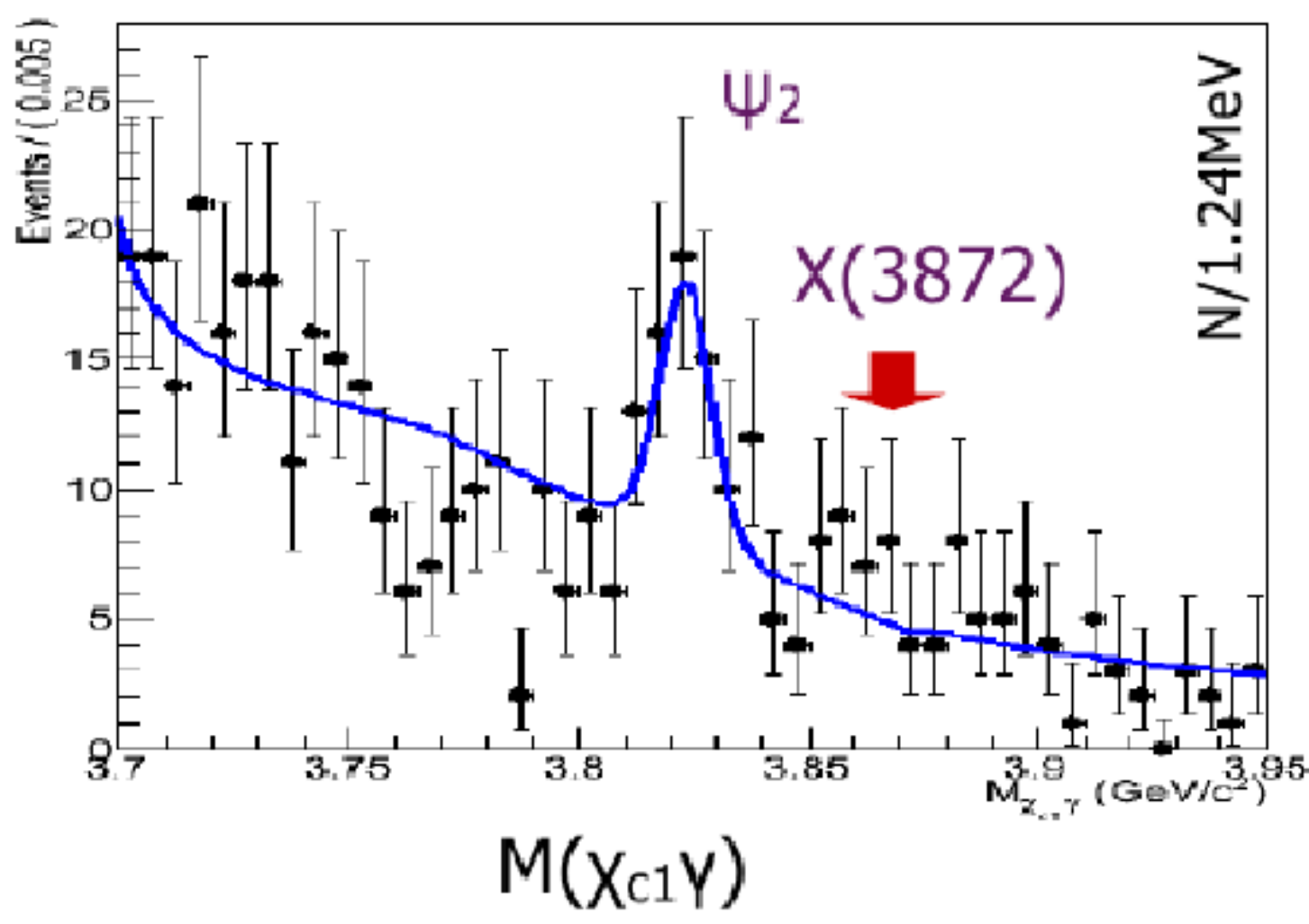}
\caption{The $M_{\gamma\chi_{c1}}$ distribution in $B$ decays from Belle.} \label{x-gchic}
\end{figure*}

In the same search for the $B\to K+\gamma\chi_{c1}$ transition, Belle found evidence of a
structure around $3.82~\gevcs$ for the first time; this is in good agreement with the  potential-model
prediction for the $\psi_2$ excited charmonium state~\cite{psi2-theo}. Figure~\ref{x-gchic} shows Belle's $M_{\gamma\chi_{c1}}$ distribution.
The significance of the $\psi_2$ signal is $4.2\sigma$ including the systematic error; the fitted width of this state is $\Gamma(\psi_2) = 4\pm6~\mev$.

\section{$X(3915)\to\omega\jpsi$}

\begin{figure*}[htbp]
\centering
\includegraphics[height=4.1cm]{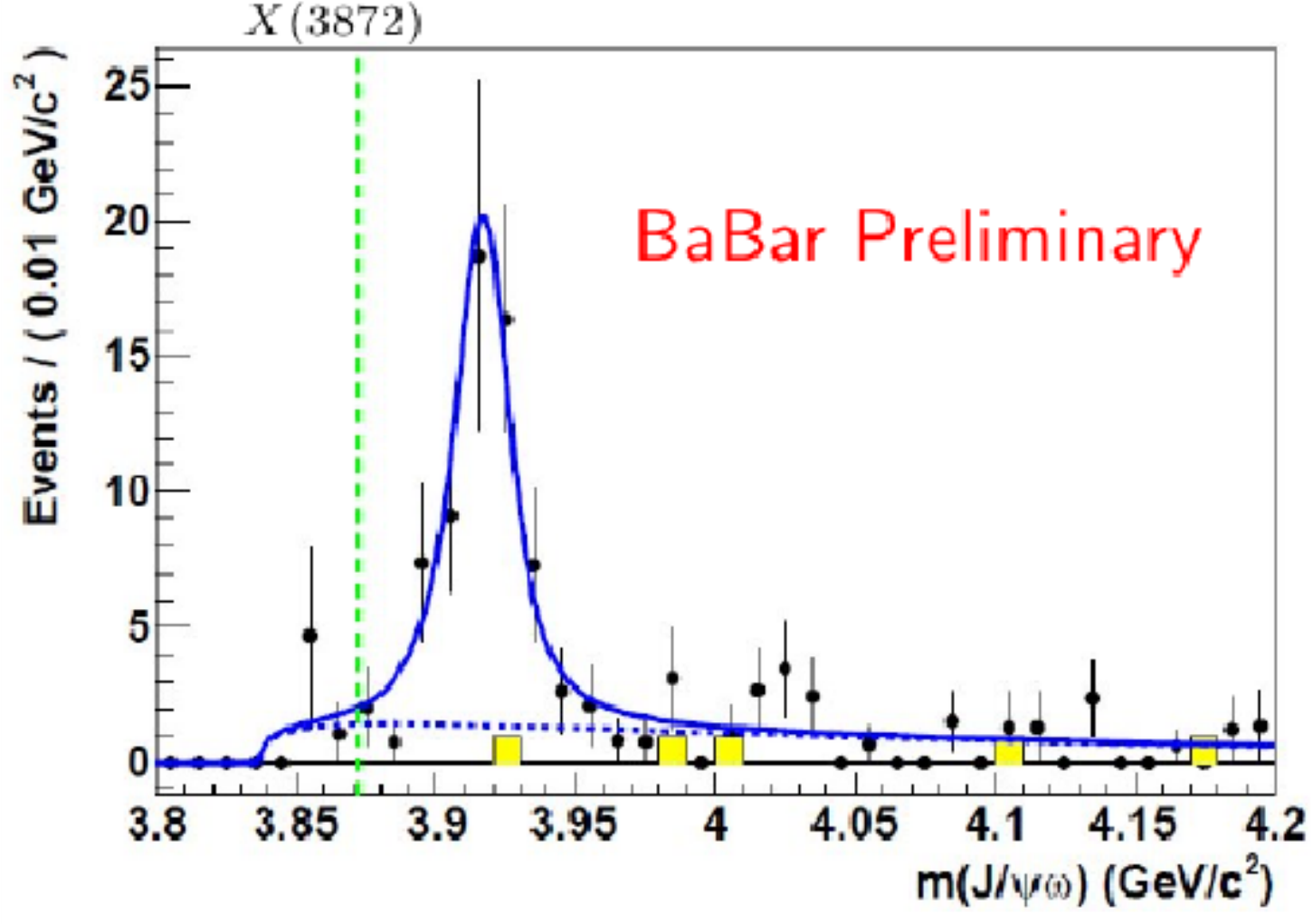}
\caption{The $M_{\omega\jpsi}$ distribution in $\gamma\gamma$ collision
from  BaBar. } \label{x3915}
\end{figure*}

The $X(3915)$ state was observed by Belle for the first time~\cite{x3915-belle} in the search for the $\omega\jpsi$
final state in $\gamma\gamma$ collisions. Belle measured a mass and width of $M_{X(3915)} = 3914\pm3\pm2~\mevcs$ and
$\Gamma_{X(3915)} = 23\pm10^{+2}_{-8}~\mev$. The $X(3915)$ is a good candidate of a
$C-$even excited charmonium state. BaBar confirms the signal in the same
process with its $519~\infb$ data sample.
From their measurement, illustrated in Fig.~\ref{x3915}, BaBar obtains a mass and width of $M_{X(3915)}=3919.4\pm2.2\pm1.6~\mevcs$ and
$\Gamma_{X(3915)}=13\pm6\pm3~\mev$. Assuming a spin of $J=0$ for the $X(3915)$, BaBar gets
$\Gamma_{\gamma}\times\BR(\omega\jpsi)=52\pm10\pm3~\ev$; alternatively, assuming
$J=2$, BaBar gets $\Gamma_{\gamma}\times\BR(\omega\jpsi) = 10.5\pm1.9\pm0.6~\ev$.

\section{Study on $\gamma\gamma\to\pp\eta_{c}$}

\begin{figure*}[htbp]
\centering
\includegraphics[height=4.cm]{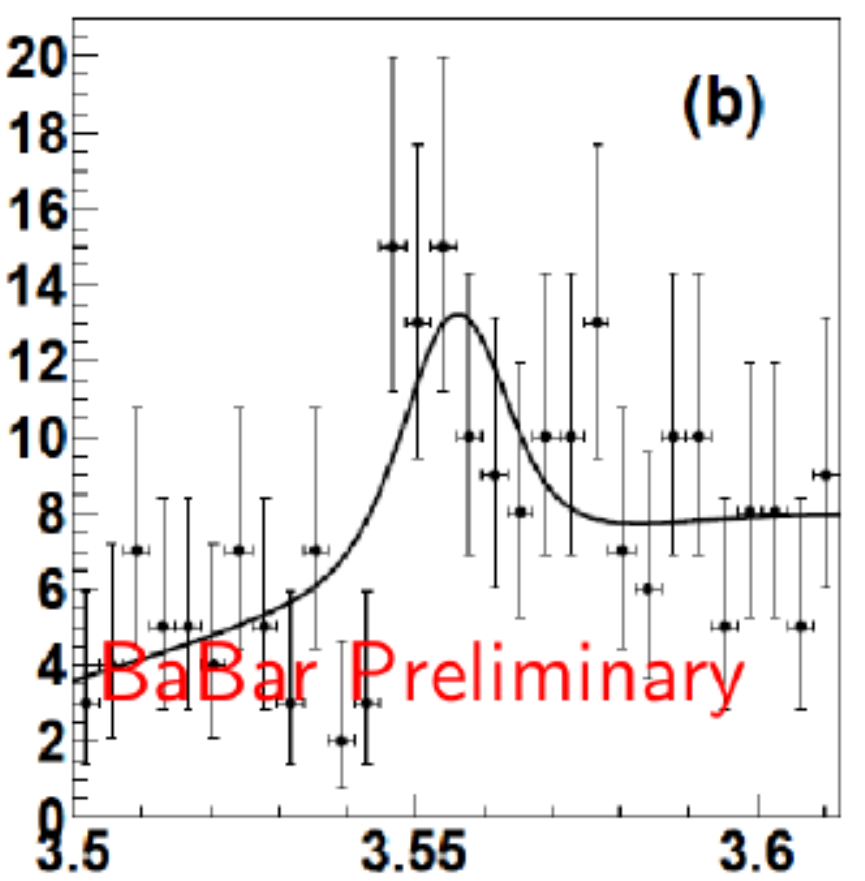}
\includegraphics[height=4.cm]{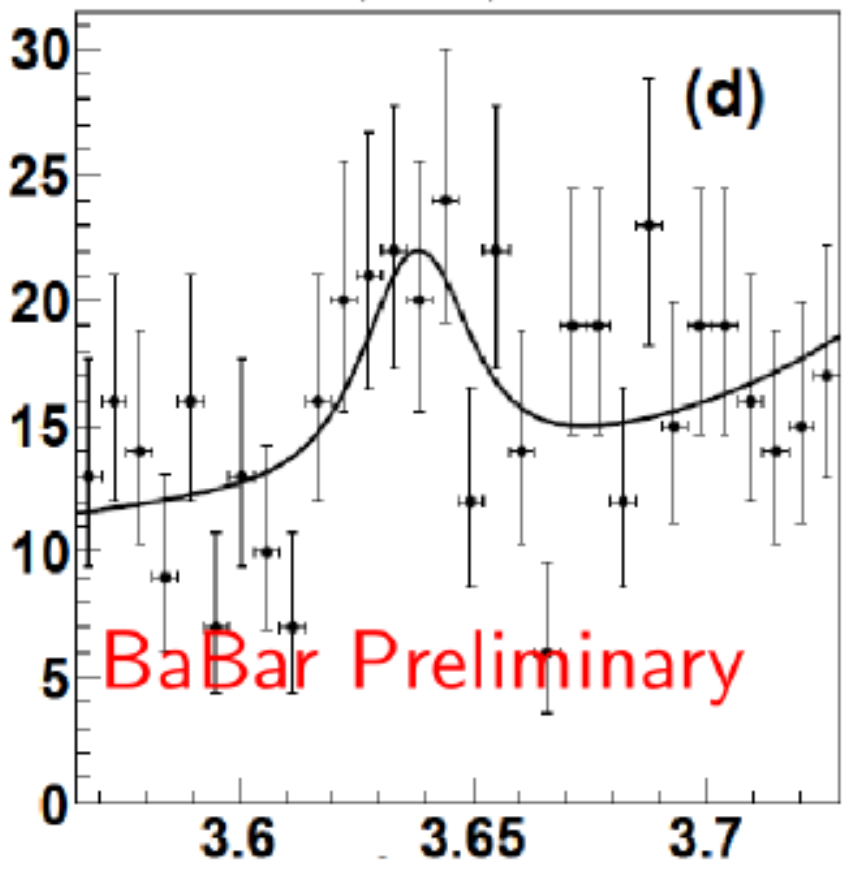}
\includegraphics[height=4.cm]{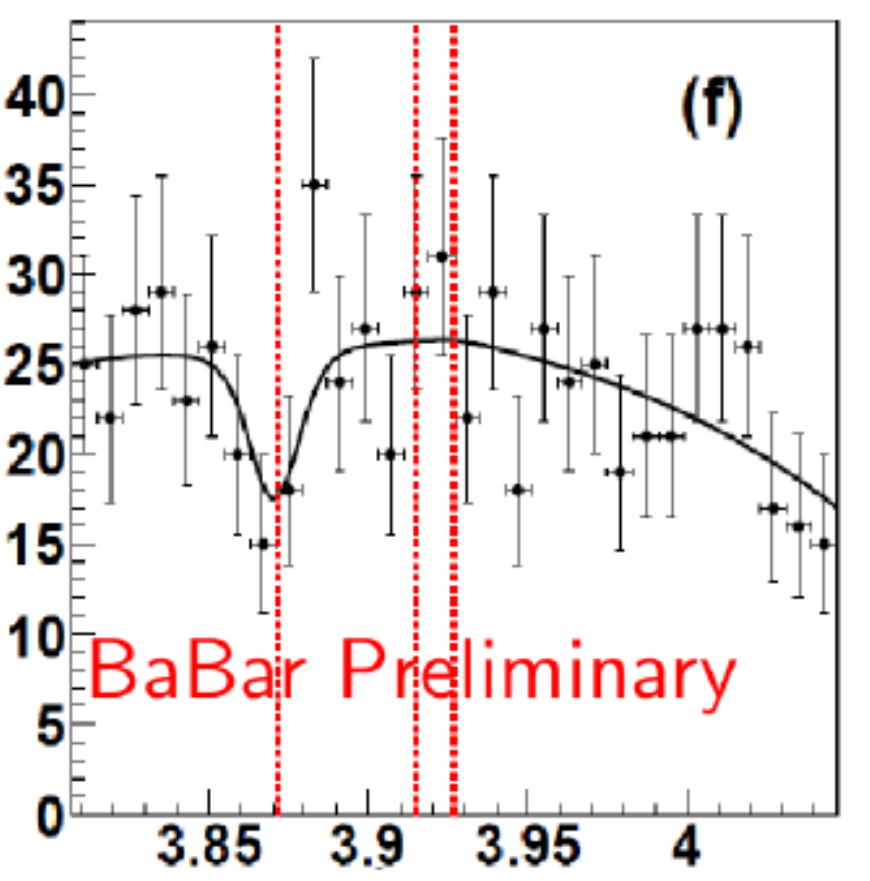}
\caption{The $M_{\pp\eta_c}$ distributions in $\gamma\gamma$ collisions from BaBar.
There are obvious $\chi_{c2}$ and $\eta_c(2S)$ signals, but no evidence for $\x$,
$X(3915)$ nor $\chi_{c2}(2P)$.} \label{ppetac}
\end{figure*}

BaBar examined the $\pp\eta_c$ final state in $\gamma\gamma$ collisions with $473.9~\infb$ data
to study the $C-$even charmonium states.
Their $M_{\pp\eta_c}$ distributions in different ranges are shown in Fig.~\ref{ppetac}.
The $\chi_{c2}$ and $\eta_c(2S)$ states are observed, but there is no evidence for
 $\x$, $X(3915)$ nor $\chi_{c2}(2P)$. The branching fractions
are determined to be $\frac{\BR(\chi_{c2}(1P)\to\eta_c\pp)}{\BR(\chi_{c2}(1P)\to\kkpi)} = 14.5^{+10.9}_{-8.9}\pm7.3\pm2.5$ and
$\frac{\eta_c(2S)\to\eta_c\pp}{\eta_c(2S)\to\kkpi} = 4.9^{+3.5}_{-3.3}\pm1.3\pm0.8$ for the observed states;  the upper limit
$\Gamma_{\gamma\gamma}(X)\cdot\BR(X\to\eta_c\pp) < 11.1/16/19~\ev$ for $\x/X(3915)/\chi_{c2}(2P)$ at 90\% C.L. for the other states.
is obtained.

\section{Studies via Initial State Radiation}

BaBar and Belle have reported studies of several exclusive initial state radiation (ISR) processes.
BaBar updates their prior scans of $\EE\to\ppjpsi$~\cite{babar-ppjpsi} and $\EE\to\pp\psp$, while Belle performs a new scan of
$\EE\to\eta\jpsi$.

\begin{figure*}[htbp]
\centering
\includegraphics[height=4.cm]{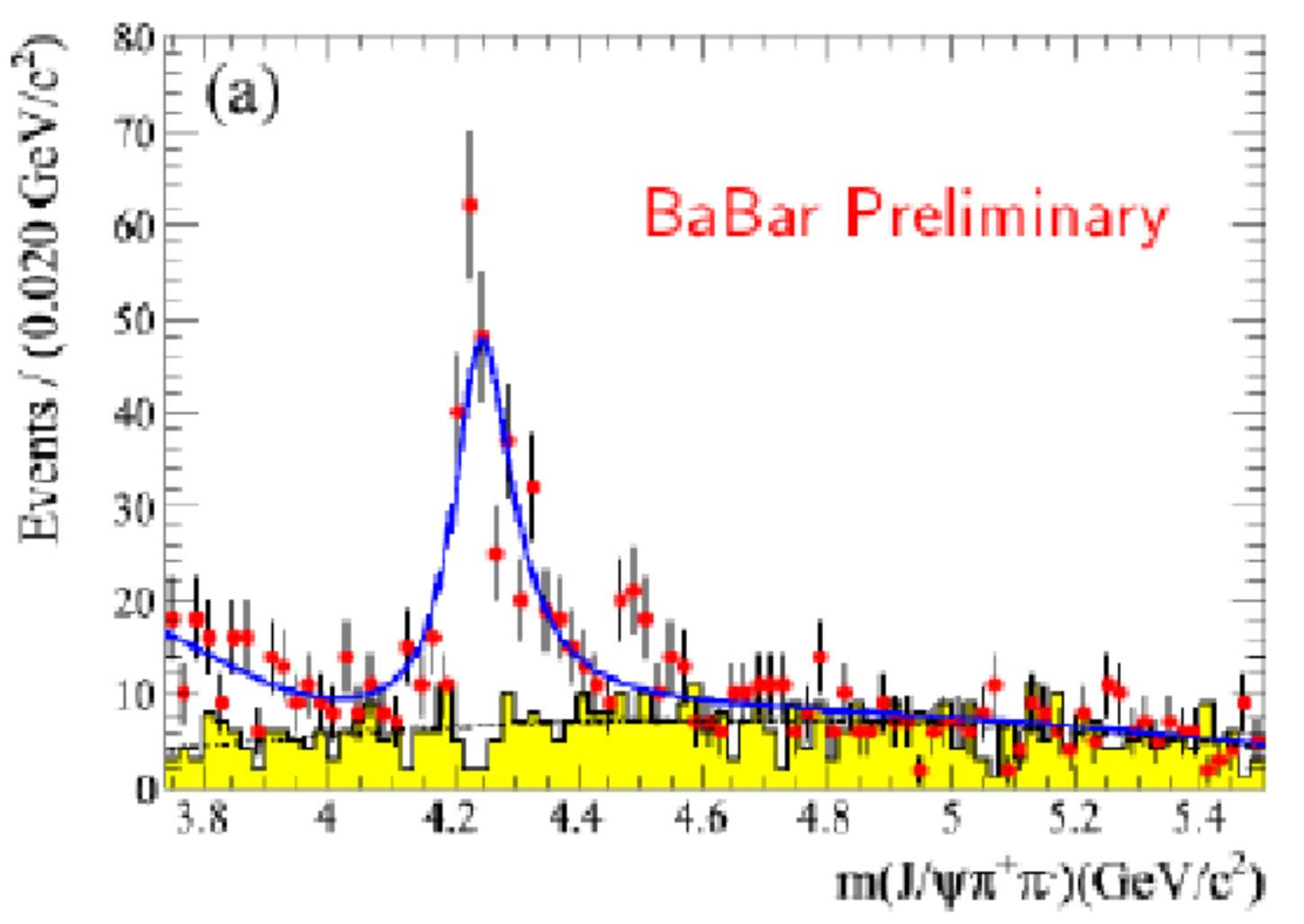}
\includegraphics[height=4.cm]{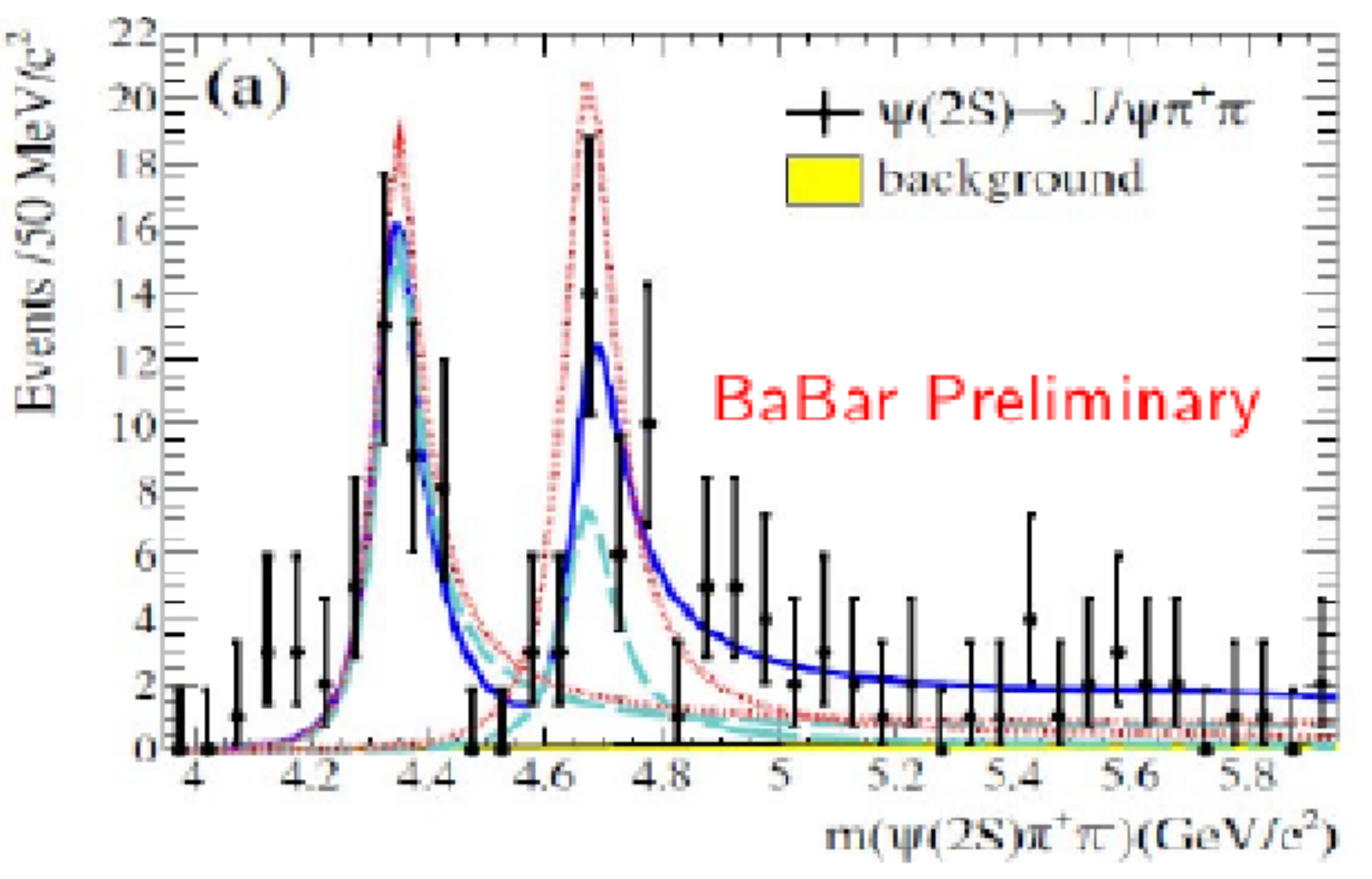}
\caption{The $M_{\pp\jpsi}$ distribution and $M_{\pp\psp}$ distribution from BaBar.}
\label{isr-babar}
\end{figure*}

Figure~\ref{isr-babar} shows the results from BaBar.
The broad structure $Y(4008)$~\cite{belle-ppjpsi} seen earlier by Belle is not confirmed; instead, the
 $\pp\jpsi$ events below $4~\gevcs$ are considered to be a
contribution from the tail of $\psp\to\pp\jpsi$. In $\EE\to\pp\psp$, BaBar confirms the $Y(4660)$~\cite{belle-pppsp}
structure for the first time. The measurement from BaBar gives $M_{Y(4660)} = 4669\pm21\pm10~\mevcs$
and $\Gamma_{Y(4660)} = 104\pm48\pm10~\mev$.

\begin{figure}[htbp]
\centering
\includegraphics[height=4.3cm]{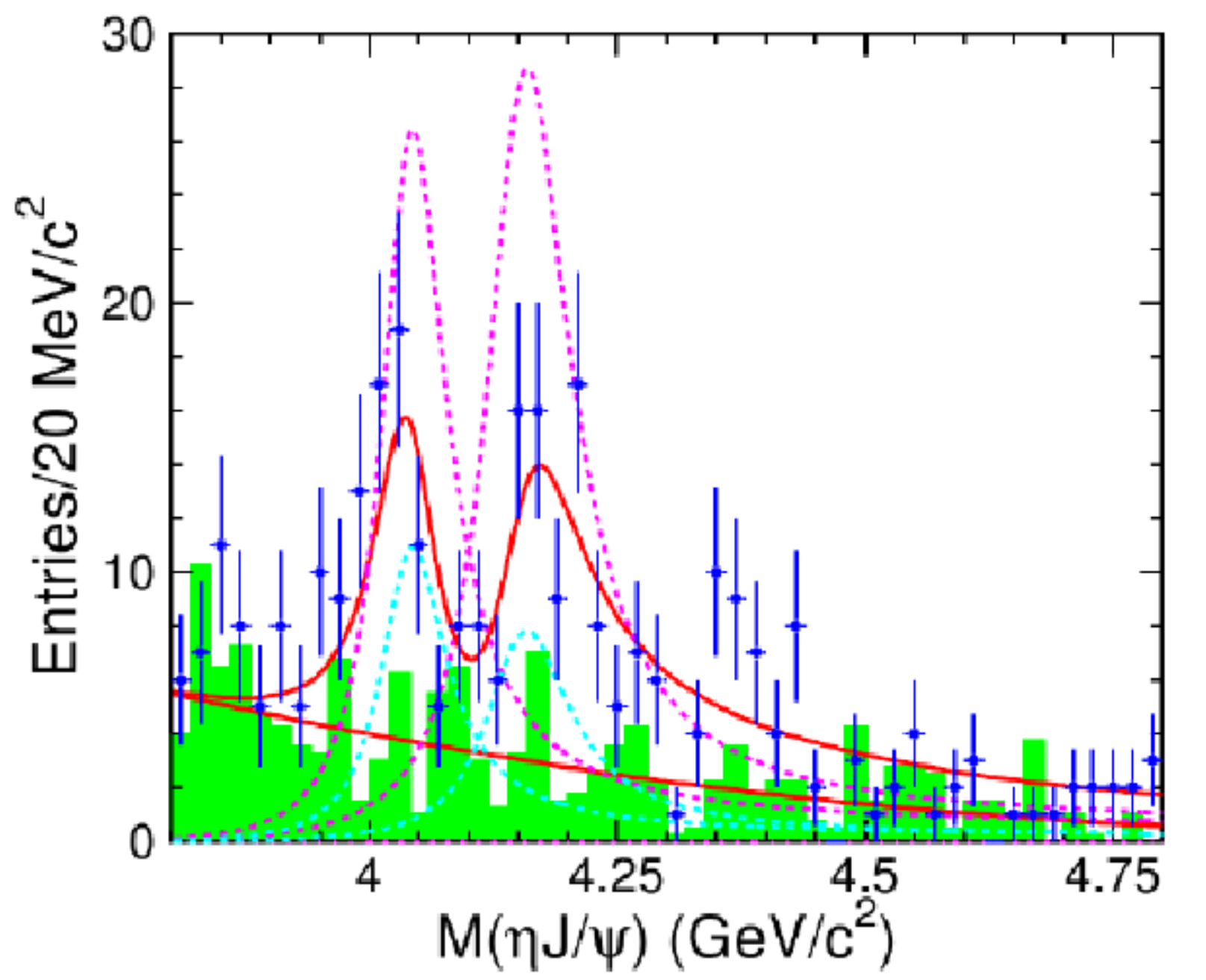}
\caption{The $M_{\eta\jpsi}$ distribution from ISR.}
\label{isr-belle}
\end{figure}

Using a $980~\infb$ data sample, Belle searches for $\EE\to\eta\jpsi$ via ISR for the first time.
There are distinct $\psi(4040)$ and
$\psi(4160)$ states observed but no $Y$ state observed in this final state.
This is also the first observation of $\psi(4040)$ and
$\psi(4160)$ decays to a final state without charmed pairs. Figure~\ref{isr-belle} shows the $M_{\eta\jpsi}$
distribution. There are two solutions with the same
goodness of the fit when fitting the $M_{\eta\jpsi}$ spectrum with coherent $\psi(4040)$ and $\psi(4160)$ resonances.
The branching fractions are measured to be $\BR(\psi(4040)\to\eta\jpsi) = (0.59\pm0.11\pm0.14)\%$ and
$\BR(\psi(4160)\to\eta\jpsi) = (0.50\pm0.07\pm0.11)\%$  for one solution, and
$\BR(\psi(4040)\to\eta\jpsi) = (1.44\pm0.18\pm0.18)\%$ and $\BR(\psi(4160)\to\eta\jpsi) = (1.83\pm0.21\pm0.24)\%$
for the alternate solution. These unusually large branching fractions at the 1\% level correspond to partial
widths of roughly $1~\mev$.

\section{Summary}

There are important experimental results on new and conventional charmonium states from BaBar and Belle. The measurement
on $\x\to\ppjpsi$ has been updated, and possible $C-$odd partner and charged partner of $\x$ are searched for but not observed.
The $\psi_2$ state is observed for the first time. The $X(3915)$ and $Y(4660)$ are confirmed for the first time. The 
$\EE\to\eta\jpsi$ final state has been
scanned via ISR, and $\psi(4040)$ and $\psi(4160)$ states but not the $Y$ state are observed.



\end{document}